\documentclass[prd,aps,floats,twocolumn,twoside,preprintnumbers,
superscriptaddress,floatfix,nofootinbib,showpacs]{revtex4}
\usepackage{graphicx,pstricks,rotating,amsmath}
\usepackage{amssymb,bbm}
\usepackage{slashed}
\begin{document}


\title{Inferring type and scale of noncommutativity from the PTOLEMY experiment}

\author{Raul Horvat} 
\affiliation{Ruder Boskovic Institute, Division of Experimental Physics, Bijenicka 54 Zagreb, Croatia}
\email[]{raul.horvat@irb.hr}
\author{Josip Trampetic}
\affiliation{Ruder Boskovic Institute, Division of Experimental Physics, Bijenicka 54 Zagreb, Croatia}
\email[]{josip.trampetic@irb.hr}
\address{Max-Planck-Institut f\"ur Physik, (Werner-Heisenberg-Institut), F\"ohringer Ring 6, D-80805 M\"unchen, Germany}
\email[]{trampeti@mpp.mpg.de}
\author{Jiangyang You}
\affiliation{Rujer Boskovic Institute, Division of Physical Chemistry, 
Bijenicka 54 Zagreb, Croatia}
\email[]{jiangyang.you@irb.hr}

\date{\today}

\begin{abstract}
If neutrinos are Dirac particles and their right-handed components can be copiously produced in the early universe, then they could influence  a direct observation of
the cosmic neutrino background, which, most likely, will come about with the
recently proposed PTOLEMY experiment.  For coupling of photons to the right-handed neutrinos we use a state-of-the-art version of gauge field theory
deformed by the spacetime noncommutativity, to disclose by it not only the decoupling temperature for the said neutrino component, but also the otherwise hidden coupling temperature.  Considering two relevant processes, the plasmon
decay and the neutrino elastic scattering, we study the interplay between
the structure of the noncommutativity parameter $\theta^{\mu \nu}$ (type of
noncommutativity) and the reheating temperature after inflation to obtain
otherwise elusive upper bound on the scale of noncommutativity $\Lambda_{\rm NC}$. If  PTOLEMY enhanced capture rate is due to spacetime noncommutativity, we verify that a nontrivial maximum upper bound on $\Lambda_{\rm NC}$ (way below the Planck scale) emerges for a space-like $\theta^{\mu \nu}$ and  sufficiently high reheating temperature.    
\end{abstract}




\maketitle

While by means of constantly improving direct  detection techniques, the cosmic microwave photon background (CMB) has provided us with a great deal of many cosmological parameters, the undisputed existence of a cosmic neutrino background (C$\nu$B) has not been hitherto directly proven, in spite of the role cosmic neutrinos had played in the evolution and the structure of the cosmos \cite{Olive:2016xmw}. Cosmic neutrinos, whose relic background is today in the form of a non-relativistic gas of particles, have been  directly related to the Big Bang Nucleosynthesis (BBN) \cite{Sarkar:1996,Kolb}, and still, after neutrinos intrinsically have been shown to have a rest mass, some percentage of dark matter of the universe is composed of them. The possibility to directly detect the present-day cold sea of relic neutrinos is about to come with the recently proposed PTOLEMY experiment \cite{Betts:2013uya}.

A first pertinent proposal to detect such a cold sea of neutrinos at the present day temperature of $T_{\nu} \approx 2$ K dates back from 1962, when a no-threshold process, the neutrino capture on tritium  $\nu_{e} \,+\, ^3$H$\:\rightarrow {^3}$He$ \;+\; e^{-}$,  
was put forward by Weinberg \cite{Weinberg:1962zza}. The first attempt to make use of the process for experimental use was given in Ref. \cite{Irvine:1983nr}, followed by a bunch of other attempts which all have shown futile \cite{Langacker:1982ih,Duda:2001hd,Cocco:2007za}. And finally, the PTOLEMY experiment has been proposed \cite{Betts:2013uya}, with  the energy resolution for the final state electrons in the ballpark of the present neutrino mass bounds, a necessary  prerequisite for a successful detection.

Recently, the authors of Ref. \cite{Horvat:2017gfm} have analyzed how the thermal production and subsequent decoupling of right-handed neutrinos $\nu_{R}$ in the early universe can influence the capture rate of right-helicity neutrinos $\nu_{r}$ in the PTOLEMY experiment.  Analysis  pursued along the similar
lines can be found also in \cite{Zhang:2015wua,Huang:2016qmh}. In the focus of \cite{Horvat:2017gfm} was a gauge field theoretical model incorporating spacetime noncommutativity (NC), which had been previously shaped next to the mature phase, so to ultimately display a trouble-free UV/IR behavior at the quantum level, both without and with supersymmetry \cite{Schupp:2008fs,Trampetic:2015zma,Horvat:2011iv,Horvat:2011bs,Horvat:2011qg,Horvat:2013rga,Horvat:2015aca,Martin:2016zon,Martin:2016hji,Martin:2016saw}. Two salient features of the model, both of which being relevant for C$\nu$B, is the Seiberg-Witten map based \cite{Seiberg:1999vs,Madore:2000en,Calmet:2001na,Aschieri:2002mc,Schupp:2002up,Minkowski:2003jg} $\theta$-exact formulation of NCQFT \cite{Horvat:2011iv,Horvat:2011qn,Horvat:2012vn}, and a tree-level vector-like coupling between neutrinos and photons \cite{Schupp:2002up,Minkowski:2003jg}, with that latter being responsible for the copious production of $\nu_{R}$s in the early universe.   

Assuming neutrinos to be Dirac particles, it has been shown \cite{Zhang:2015wua,Horvat:2017gfm} that the PTOLEMY capture rate can be at most increased up to around 20\%, if neutrinos are produced thermally. For nonthermal production, see \cite{Huang:2016qmh}. The physics behind this estimate rests on the latest bound \cite{Ade:2015gva} on the effective number of neutrino species  $N_{eff}$ and the fact that for propagating neutrinos it is their helicity that is conserved \cite{Duda:2001hd}. Namely, while at birth and at freeze out $\nu_{R}$s practically coincide with $\nu_{r}$s since neutrinos are then ultra-relativistic, the C$\nu$B is non-relativistic today and therefore $\nu_{r}$s are captured in the process equally likely as their left-helical partners $\nu_{l}$s do. This was discussed at length in \cite{Long:2014zva} where also the most accurate expression for the capture rate was given.

For coupling of photons to the right-handed neutrinos we use a state-of-the-art version of gauge field theory (GFT) deformed by the spacetime noncommutativity  (NC) \cite{Horvat:2011iv,Horvat:2011bs,Horvat:2011qg,Horvat:2013rga,Horvat:2015aca,Martin:2016zon,Martin:2016hji,Martin:2016saw} employed in our previous work \cite{Horvat:2017gfm}. Namely, generically electrically neutral matter fields will be promoted via (hybrid) Seiberg-Witten (SW) maps \cite{Horvat:2011qn} to noncommutative fields that couple via star commutator to photons and transform in the adjoint representation of U(1). The inclusion of all gauge covariant coupling terms is furthermore a prerequisite  for reasonable UV behavior. Taking this into account we arrive at the following model of the SW type NC  U(1) GFT \footnote{ Here we set the coupling constant $e=1$. To restore the coupling constant one simply substitute $A_\mu$ by $eA_\mu$, then divide the gauge field Lagrangian by $e^2$. Such a model emerged also in a direct construction of the Moyal deformed standard model~\cite{Chaichian:2001py}, yet the explicit construction~\cite{Chaichian:2001py} includes only the left-handed neutrinos, thus inapplicable in our study here. Namely, the neutrino-photon direct vertex in the model of Ref \cite{Chaichian:2001py} is a {\it chiral} one, i.e. the only existing neutrino that appears in this interaction term is the left-handed one. More precisely, due to the hypercharge structure of the model gauge group  $\rm U_\star(3)\times U_\star(2)\times U_\star(1)$ used in \cite{Chaichian:2001py}, 
a tree-level coupling of the right-handed neutrino to the electro-magnetic field is absent. Due to the same reason the $\kappa$-value in \cite{Chaichian:2001py} is fixed to one.
} 
\begin{equation}
S=\int -\frac{1}{4}\widehat F^{\mu\nu}\star \widehat F_{\mu\nu}+i\bar{\widehat\Psi}\star
\left(\slashed{\widehat D}-m\right)\widehat\Psi
\label{swqed:action}
\end{equation}
with the NC field strength being $\widehat F_{\mu\nu}=\partial_\mu
\widehat A_\nu-\partial_\nu\widehat A_\mu-ie\kappa[\widehat A_\mu\stackrel{\star}{,}\widehat A_\nu]$ and the NC covariant derivative $\widehat D_\mu\widehat\Psi=\partial_\mu\widehat\Psi-ie\kappa[\widehat A_\mu\stackrel{\star}{,}\widehat\Psi]$, respectively. Fields $\widehat A^\mu,\widehat\Psi,....$ in particular are noncommutative fields spanned on the Moyal manifold. Here  $\widehat\Psi$ means noncommutative $\widehat\Psi_{L \choose R}$, i.e.  the NC left-right Dirac-type massive neutrino field. Coupling constant $e\kappa$ corresponds to positive multiple (or fraction) $\kappa$ of charge $|e|$. The Moyal $\star$-product above is associative but not commutative - otherwise the proposed coupling to the noncommutative gauge/photon field $\widehat A_\mu$ would of course be zero.

All the fields in this action are images under hybrid Seiberg-Witten maps \cite{Horvat:2011qn,Martin:2015nna} of the corresponding commutative fields $A_\mu$ and $\Psi$. In the original SW work and in virtually all subsequent applications, these maps are understood as (formal) series in powers of the noncommutativity parameter
$\theta^{\mu\nu}$. Physically, this corresponds to an expansion in momenta and is valid only for low energy phenomena. Here we shall not subscribe to this point of view and instead interpret the NC fields as valued in the enveloping algebra of  the underlying gauge group. This naturally corresponds to an expansion in powers of  gauge field $A_\mu$  and hence in powers of the coupling constant. 


Noncommutative fields $(\widehat A^\mu,\widehat \Psi,....)$ from the action (\ref{swqed:action}) are functionals of commutative fields ($A_\mu,\Psi,....$) expanded in powers of ordinary gauge field $A_\mu$ via $\theta$-exact SW maps
\begin{equation}
\begin{split}
\widehat A_\mu&=A_\mu-\frac{e\kappa}{2}\theta^{ij}A_i\star_2(\partial_jA_\mu+F_{j\mu})+\mathcal O(A^3)\,,
\\
\widehat\Psi&=\Psi-e\kappa\theta^{ij}A_i\star_2\partial_j\Psi+\mathcal O(A^2)\Psi\,,
\end{split}
\label{exactPsi}
\end{equation}
where the $\star_2$-product, defined in many papers as in \cite{Horvat:2017gfm}, is commutative, but not associative. For the SW map of the gauge field strength $\widehat F^{\mu\nu}$ see \cite{Trampetic:2015zma}. Here $\Psi$ means commutative $\Psi_{L \choose R}$, i.e.  left-right Dirac-type\footnote{Note that instead of SW map of Dirac neutrinos $\Psi$ one may consider a {\it chiral} SW map, which is compatible with grand unified models having chiral fermion multiplets \cite{Aschieri:2002mc}.} massive neutrino field. However further on we only consider the right-handed neutrinos to be directly (tree-level) coupled to photons via NC mechanism, as a new contribution not present in the SM.

By using $\theta$-exact SW maps (\ref{exactPsi}) to express the action \eqref{swqed:action} in terms of commutative fields, we find the following $\theta$-exact Lagrangian up to the first order in gauge fields $A_\mu$, and for the diagonal in flavor (flavor blind) interaction:
\begin{equation}
\begin{split}
{\cal L}&=\bar\Psi\Big[\slashed A\stackrel{\star}{,}\Psi\Big]
-e\kappa\Big(\theta^{\mu\nu}A_\mu
\star_2 \partial_\nu\bar\Psi\Big)\Big(i\slashed\partial - m\Big)\Psi
\\&\quad
-e\kappa\bar\Psi
\Big(i\slashed\partial - m\Big)\Big(\theta^{\mu\nu}
A_\mu\star_2\partial_\nu\Psi\Big)+\bar\Psi\mathcal{O}\big(A^2\big)\Psi\,.
\label{pf:int}
\end{split}
\end{equation}
From (\ref{pf:int}) we finally obtain the following Feynman rules for the right-handed Dirac neutrinos 
\begin{eqnarray}
&&\Gamma_{\mu}(p,q)\Big|_R= ie\kappa \frac{1}{2} (1 + \gamma_5)
\nonumber\\
&&\cdot\Big[(/\!\!\! p - m) {(\theta q)}_{\mu}-(p\theta q)\gamma_{\mu} -{(\theta p)}_{\mu}{/\!\!\! q}\Big]F(p,q).
\label{1}
\end{eqnarray}
Here the function $F(q, p)$ is given by,
\begin{equation}
F(q,p)=\frac{\sin \frac{q\theta p}{2}}{\frac{q\theta p}{2}},\;\;q\theta p \equiv q_\alpha \theta^{\alpha\beta} p_\beta\,,
\label{2}
\end{equation}
and $\theta^{\alpha\beta}$ is the NC parameter, usually given by
$c^{\alpha\beta}/\Lambda^2_{\rm NC}$, where $\Lambda_{\rm NC}$ is the scale of noncommutativity and $c^{\alpha\beta}$ are numbers of order one. 

An important note about the time-, space- and light-like NC QFT's  is in order.
It was shown \cite{Connes:1997cr,Douglas:1997fm,Seiberg:1999vs} that fields theories with space-like noncommutativity arise from a decoupling limit of string theory involving D-branes with non-zero space-like NS-NS B fields. In this case all string modes decouple and one is left with a unitary field theory. On the other hand, field theories with time-like noncommutativity were shown not to be unitary \cite{Gomis:2000zz} since a decoupled field theory limit for D-branes with a time-like B field does not exist \cite{Seiberg:2000ms,Gopakumar:2000na,Barbon:2000sg}. Besides, such theories exhibit noncausal behavior \cite{Seiberg:2000gc,AlvarezGaume:2000bv}. Finally, in spite of the nonlocality in the time coordinate due to  $\theta^{0i} \neq 0$, quantum theories with light-like noncommutativity were shown to be unitary \cite{Aharony:2000gz} putting them on equal footing with those more common theories with space-like noncommutativity.

In Ref. \cite{Horvat:2017gfm} $\nu_{R}$s are produced in the early universe via the plasmon decay, $\gamma_{pl.} \rightarrow \bar{\nu}_R \nu_R$ \footnote{In \cite{Horvat:2017gfm} it has been  shown that the plasmon decay rate contain only $\theta^{0i}$ (i=1, 2, 3) parts of the NC parameter $\theta^{\mu \nu}$. As a consequence, the plasmon decay rate is nonzero for light-like noncommutativity only.}, enabled by the tree-level vector-like coupling (\ref{1}) between photons and neutrinos $\bar\nu_R\nu_R\gamma$  in the noncommutative scenario.\footnote{Since other properties of noncommutative theory producing vertex (\ref{1}) with the $\kappa$ parameter was in detail discussed in \cite{Horvat:2017gfm}, there is no need to repeat the same discussion here. Nevertheless appearances of a universal $\kappa$-value across all flavor generations as in~\cite{Horvat:2011qn} actually allows most general neutrino mixing in the gauge invariant mass/Yukawa term constructions. Hence in the rest of the paper we will deal with universal but otherwise arbitrary $\kappa$ parameter. }

When comparing the plasmon decay rate with the Hubble expansion
parameter, a distinctive feature shows up in the numerical plot of
$\Lambda_{\rm NC}$ versus decoupling temperature $T_{dec}$ - a coupling
temperature (see Fig. 2 in \cite{Horvat:2017gfm}). The coupling temperature $T_{couple}$ shows up when the reheating temperature after inflation $T_{reh}$ is high enough, and designate the temperature when, during cooling after the Big Bang, $\nu_{R}$s first time enter thermal equilibrium with the rest of the universe. After spending a while in thermal equilibrium, $\nu_{R}$s decouple again at $T_{dec}$. Hence, if $T_{reh} > T_{couple}$, $\nu_{R}$s stay in thermal equilibrium in the temperature range between $T_{couple}$ and $T_{dec}$. In turn, this translates into an exceptional maximum upper bound on $\Lambda_{\rm NC}$, of order of $10^{-4} M_{Pl}$. If, on the other hand, $T_{reh} < T_{couple}$, the coupling temperature ceases to exist  and the upper bound on $\Lambda_{\rm NC}$ inferred from the experiment does substantially depend on $T_{reh}$, being always less than the exceptional one.  Coupling temperature $T_{couple}$ inferred from the plasmon decay is as high as $\rm{10^{15}}$ GeV, being of the same order as the maximum reheating temperature considering perturbatively decaying inflaton \cite{Maity:2017thw}, and therefore the exceptional bound on $\Lambda_{\rm NC}$ (independent of $T_{reh}$) may no longer exists. 

In the present paper we aim to reassess the scenario and re-derive the bounds on  
$\Lambda_{\rm NC}$ by inclusion of the yet another process, the elastic right-handed neutrino scattering on electrons $e\nu_R  \rightarrow e\nu_R $.
   


Additional motivation to use scattering mechanisms is the issue of the unitarity test of NCQFT. Namely plasmon decay rate is nonzero only in theories with light-like noncommutativity where for example $\theta^{0i} = -\theta^{1i},\;\,\forall i=1,2,3$  \cite{Aharony:2000gz}, while it vanishes identically for the space-like type of noncommutativity, $\theta^{0i} = 0$, $\theta^{ij} \not= 0$. Furthermore the time-like noncommutativity, $\theta^{0i} \neq 0$; $\theta^{ij} =0, \;\forall i,j=1,2,3,$  does not lead to unitary quantum field theories, and therefore will not be considered here.

The scattering process $e\nu_R \to e\nu_R$ takes place in the t-channel which is topologically different from the plasmon decay (essentially the s-channel process). In turn, this entails that the scattering process proceeds equally well for both the space-like and light-like type of noncommutativity. See also \cite{Horvat:2010sr} where the role of neutrino scattering in the ultra-high energy cosmic ray experiments was highlighted.


We show here that with a more familiar and appealing space-like noncommutativity, the elastic neutrino--electron scattering would lessen $T_{couple}$ as well as the belonging bounds on $\Lambda_{\rm NC}$ by a few orders of magnitude, that means, safely below the maximum reheating temperature. This way, the exceptional upper bound on $\Lambda_{\rm NC}$ (shown to be somewhere halfway between the weak and the Planck scale) can be able to survive, and ultimately can be drawn out  from the PTOLEMY experiment.

From the Feynman rule (\ref{1}) and by using a standard techniques, one obtains the total cross section of the 
$\nu_R e \rightarrow \nu_R e$ scattering from the lowest order NC t-channel amplitude (photon exchange diagram), i.e. diagram where $\bar\nu_R\nu_R\gamma$ vertex is the noncommutative one in the model (\ref{pf:int}), while $\bar e e\gamma$ vertex is the SM one, giving
\begin{eqnarray}
&&\hspace{-1.4cm}\sigma_{\rm NC}(e\nu\to e\nu)=\frac{\kappa^2\alpha^2}{16E^2}
\int \sin\vartheta d\vartheta\frac{4+(1+\cos\vartheta)^2}{(1-\cos\vartheta)^2}
\;\cal I,
\label{3}\\
\cal I&=&\int d\varphi \;4\sin^2\frac{p\theta p'}{2}=2\Big(1-\cos(\xi)J_0(\zeta)\Big),
\label{4}\\
\xi&=&\frac{E^2}{\Lambda^2_{\rm NC}}c_{03}\big(\cos\vartheta-1\big),\;
\nonumber\\
\zeta&=&\frac{E^2}{\Lambda^2_{\rm NC}}
\sin\vartheta\Big({\rm sign}(c_{01}-c_{03})\Big)
\nonumber\\
&&\cdot\sqrt{(c_{01}-c_{13})^2+(c_{02}-c_{23})^2} .
\label{5}
\end{eqnarray}
As we expected there is in (\ref{3}--\ref{5}) a presence of both the time-like and the space-like components of $\theta^{\mu \nu}$, respectively. Thus, the scattering rate is in principle nonzero for both types of noncommutativity. More specifically, for space-like noncommutativity $\theta^{0i} = 0$, the relevant factor in $\zeta$ boils down to $\sqrt{c_{13}^2 + c_{23}^2}$. On the other hand, for the light-like noncommutativity the factor in $\zeta$ is reduced to $\sqrt{c_{02}^2 + c_{03}^2}(=\sqrt{c_{12}^2 + c_{13}^2})$.  

In what follows we shall deal only with the space-like type of noncommutativity, the type having the most elegant embedding in string theory, and study the copious production of $\nu_R$s in the early universe using our fully-fledged NC model. This then singles out the scattering process as a dominant one for $\nu_R$ production.  

In order to secure the integration over $\vartheta$ in (\ref{3}),  
we use a Debye mass $m_D = \sqrt{g_{*}^{ch}} T/3$ as a regulator, 
where $g_{*}^{ch}$ counts the total charged (effectively massless) degrees of freedom. This way we end up with the expression for the total cross section as
\begin{figure}[t]
\begin{center}
\includegraphics[width=6cm,angle=0]{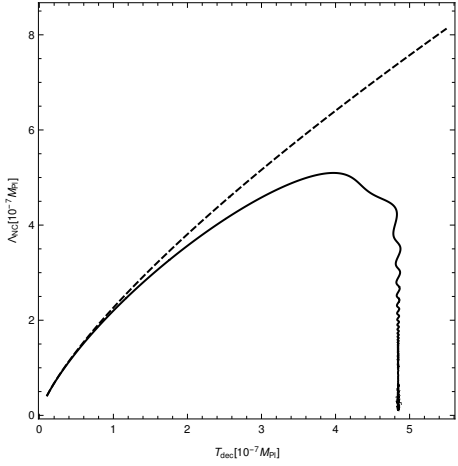}
\end{center}
\caption{Numerical plot of the scale of noncommutativity $\Lambda_{\rm NC}$ versus ``decoupling temperature'' $T_{dec}$ evaluated from (\ref{8}) for the full-$\theta$ contribution to the elastic $e\nu_R\to e\nu_R$ scattering amplitude (solid curve) and its first-order in $\theta$ approximation  (dashed curve), for  $g_{*}\simeq g_{*}^{ch}\simeq100$ and $\kappa=1$. }
\label{fig:Fig1}
\end{figure}
\begin{eqnarray}
&&\hspace{-1cm}\sigma_{\rm NC}(e\nu_R\to e\nu_R)=\frac{\kappa^2\alpha^2}{8E^2}
\int_{-1}^{1} dx\frac{4+(1+x)^2}{(1-x+\frac{\omega^2_{pl}}{E^2})^2}
\nonumber\\
&&\cdot\Bigg(1-J_0\bigg[\frac{E^2}{\Lambda^2_{\rm NC}}
(\sqrt{1-x^2})\sqrt{c_{13}^2+c_{23}^2}\bigg]\Bigg).
\label{6}
\end{eqnarray}
Comparing the rate obtained from (\ref{6}), $\Gamma_{\rm scatt.}^{\rm NC}(T_{dec})\simeq 0.18\;T_{dec}^3\,\sigma_{\rm NC}(e\nu_R\to e\nu_R)$,  with the Hubble expansion rate
\begin{equation}
\Gamma_{\rm scatt.}^{\rm NC}(T_{dec})\simeq H(T_{dec}) =
\left(\frac{8 \pi^3}{90} g_{*}(T_{dec}) \right)^{1/2} \frac{T_{dec}^2}{M_{Pl}}\,,
\label{7}
\end{equation}
where $g_{*}(T_{dec})$ counts the total number of effectively massless degrees of
freedom, we compute the following functional equation connecting  the decoupling
temperature and the scale of noncommutativity $\Lambda_{\rm NC}$
\begin{eqnarray}
&&\hspace{-1.2cm}T_{dec}\simeq1.506\times 10^{-3}
\frac{\kappa^2\alpha^2 }{ \,\sqrt{g_{*}^{ch}}}M_{Pl}
\nonumber\\
&&\hspace{-1.2cm}\cdot 
\int\limits_{-1}^1 dx\frac{4+(1+x)^2}{\big(1-x+\frac{4\pi\alpha g_{*}^{ch}}{81}\big)^2}
\Bigg(1-J_0\bigg[\frac{9T^2_{dec}}{\Lambda^2_{\rm NC}}
\sqrt{1-x^2}\bigg]\Bigg),
\label{8}
\end{eqnarray}
where $E\simeq 3T_{dec}$, and we have chosen $\sqrt{c_{13}^2 + c_{23}^2} \simeq 1$. 

Sensitivity to PTOLEMY requires small $T_{dec}$, which one can only achieve for $T^2_{dec}/\Lambda^2_{\rm NC} \ll 1$. In this limit we can use the leading order term in the Bessel function expansion: $1-J_0\Big[\frac{9T^2_{dec}}{\Lambda^2_{\rm NC}}\sqrt{1-x^2}\Big]=\frac{1}{4}\frac{81T^4_{dec}}{\Lambda^4_{\rm NC}}(1-x^2)$, and after x-integration in (\ref{8}) obtain a lower bounds on $\Lambda_{\rm NC}$ for $T_{dec}\gtrsim200$ MeV (quark-hadron phase transition), and for $T_{dec}\gtrsim200$ GeV ($EW$ phase transition), respectively. Now setting $g_* \simeq g^{\rm ch}_* \simeq 100$, and $M_{Pl}=1.221\times 10^{16}$ TeV,  a lower bounds on $\Lambda_{\rm NC}$ are:
\begin{equation}
 \Lambda_{\rm NC}  \bigg |^{T_{dec}\gtrsim200\; \rm MeV}_{T_{dec}\gtrsim200\; \rm GeV}\;\gtrsim \; \bigg |^{0.77\sqrt{\kappa}}_{137\sqrt{\kappa}}\;\; \rm TeV,
 \label{10^3}
 \end{equation}
a somewhat lower values to what we have obtained in the case of plasmon decay dominating Hubble expansion rate \cite{Horvat:2017gfm}.

Having solved (\ref{8}) numerically  we plot the solution in Fig.\ref{fig:Fig1} for $g_{*} \simeq g_{*}^{ch} \simeq 100$. Fig.\ref{fig:Fig1} shows how the nonlocality of  these field theories (featuring an explicit UV/IR mixing) may also have an important consequences for cosmology. Within the region surrounded by a solid curve, the Hubble expansion rate is always surpassed by the $\nu_R$ scattering rate. And the splitting of $T_{dec}$ into two branches, the usual decoupling temperature (a lower one) and the coupling  temperature (a higher one) is a direct consequence of UV/IR correspondence, unfolding  nicely from our full-$\theta$ NC model. Above $\Lambda_{\rm NC}^{max}$ $\nu_R$s can never attain thermal equilibrium via the NC coupling to photons and thus would have no impact on the PTOLEMY capture rate.  In contrast, with the use of the first order approximation in $\theta$ (dashed curve in Fig.\ref{fig:Fig1}), the coupling temperature is missing since the absence of the sine term in (\ref{1}) destroys the UV/IR connection. It is just the switch in the behavior of the scattering rate, from $T^5$ at low temperatures (where the full theory and the first order approximation coincide pretty accurately) to $T$ at very high temperatures, which is responsible for the closed contour in the $T_{dec}$--$\Lambda_{\rm NC}$ plane as depicted in Fig.\ref{fig:Fig1}.

From Fig.\ref{fig:Fig1} one can determine a maximum  coupling temperature to be $T_{coupl}^{max} \simeq 4.84 \times 10^{-7} M_{Pl}=5.91\times 10^{9}\; \rm TeV$, and accompanied maximum scale of noncommutativity $\Lambda_{\rm NC}^{max} \simeq 5.26 \times 10^{-7} M_{Pl}=6.42\times 10^{9}\; \rm TeV$. Those are new important bounds, being almost three orders of magnitude below those obtained from the plasmon decay in the case of  light-like type of noncommutativity \cite{Horvat:2017gfm}.

Note that the nontrivial upper bound on $\Lambda_{\rm NC}$ obtained from the plasmon decay and light-like noncommutativity may still not be there. In this case, for  $T_{dec} < 2 \times 10^{12}$ TeV the coupling temperature no longer exists, as characteristic pattern from UV/IR mixing begins to unfold beyond that temperature. Incidentally, this temperature turns out to be of the same order as the maximum reheating temperature, obtained recently in \cite{Maity:2017thw} with the assumption of the perturbative decay  of inflaton. Thus the characteristic UV/IR mixing pattern of the curve in Fig.\ref{fig:Fig1} can be wash off either by restoration of locality (e.g. by using the perturbative-in-$\theta$ expansion of the full theory) or by sufficiently low reheating temperature. On the other hand, with space-like noncommutativity the characteristic features of the curve unfolds at temperatures which are about three orders below the maximum reheating temperature, and therefore the exceptional upper bound on $\Lambda_{\rm NC}$ (independent of $T_{reh}$) could still survive. In addition, using instead the total number of effectively massless degrees of freedom relevant for MSSM ($g_{*} \simeq g_{*}^{ch} \simeq 915/4$), one can additionally reduce the coupling temperature and the bound on $\Lambda_{\rm NC}$ by about a factor of four.


Summing up, we have shown that  if  the PTOLEMY experiment would register an enhanced capture rate, then this could have  far reaching consequences for the scale and type of noncommutativity inferred from cosmology. If the space-like noncommutativity is to be realized in nature, one obtains $\Lambda_{\rm NC}^{max}$ of order of $10^{9}$ TeV, for the reheating temperature high enough. This value is consistent with the number of constraints on $\Lambda_{\rm NC}$ obtained from particle physic phenomenology  \cite{Chaichian:2001py,Hinchliffe:2002km,Ohl:2004tn,Connes:2006qv,Alboteanu:2006hh,Alboteanu:2007bp,Abel:2006wj,Alboteanu:2007by,Ettefaghi:2007zz,Horvat:2009cm,Horvat:2010sr,Horvat:2010km,Horvat:2011wh}, but still several orders below  the (theoretically appealing) string or the Planck scale. Under the same circumstances but with light-like noncommutativity, the value $\Lambda_{\rm NC}^{max}$ will strongly depend on the reheating temperature.   

\section*{Acknowledgment}
This work is supported by the Croatian Science Foundation (HRZZ) under Contract No. IP-2014-09-9582. We acknowledge the support of the COST Action MP1405  (QSPACE). J.T. would like to acknowledge support of Alexander von Humboldt-Stiftung (HRV 1028995 HFST), and  Max-Planck-Institute for Physics and W. Hollik for hospitality. J. Y. acknowledges support by the H2020 Twining project No. 692194, RBI-T-WINNING, and would also like to acknowledge the support of W.~Hollik and the Max-Planck-Institute for Physics, Munich, for hospitality.  


\end{document}